\documentclass[11pt]{article} \usepackage[dvips]{epsfig}
 \def\bb{\bibitem} \def\lb{\label}
\def\be{\begin{equation}} \def\ee{\end{equation}}
\def\ba{\begin{eqnarray}} \def\ea{\end{eqnarray}} \def\part{\partial}
\def\L{\Lambda}  \def\s{\sigma} \def\a{\alpha}
\def\ka{k\alpha} \def\tr{\tilde{r}} \def\tp{\tilde{\phi}}
\def\ts{\tilde{\s}}

\begin{document}

\begin{center}
{\bf {\Large Analytical approach to critical
collapse\\ in 2+1 dimensions}\footnote{Plenary talk presented at the
V$^{th}$ International Conference on Gravitation and Astrophysics of
Asian-Pacific Countries, Moscow 2001.}} \\
\bigskip
{\bf G\'erard Cl\'ement}$^{a}$\footnote{Email: gclement@lapp.in2p3.fr}
and {\bf Alessandro Fabbri}$^{b}$\footnote{Email: fabbria@bo.infn.it} \\ 
\medskip
{\small $^{(a)}${\it Laboratoire de  Physique Th\'eorique LAPTH (CNRS),}} \\
{\small {\it B.P.110, F-74941 Annecy-le-Vieux cedex, France}}\\
{\small $^{(b)}${\it Dipartimento di Fisica dell'Universit\`a di
Bologna and INFN sezione di Bologna,}} \\ 
{\small {\it Via Irnerio 46,  40126 Bologna, Italy}}
\end{center}

\medskip
\begin{abstract}
We present a family of time-dependent solutions to 2+1 gravity
with negative cosmological constant and a massless scalar field as
source. These solutions are continuously self-similar near the central
singularity. We analyze linear perturbations of these solutions, and
discuss the subtle question of boundary conditions. We find two
growing modes, one of which corresponds to the linearization of static
singular solutions, while the other describes black hole formation. 

\end{abstract}
\bigskip

\section{Introduction}

In recent numerical simulations \cite{CP} of gravitational collapse of
a massless scalar field in 2+1 dimensions with negative cosmological
constant $\Lambda$, threshold solutions were observed, exhibiting
features  characteristic of critical collapse \cite{chop}, namely
power-law scaling and continuous self-similarity (CSS). It has been
argued by Garfinkle \cite{gar} that this behavior is governed by an
exact CSS solution of the $\Lambda = 0$ theory. To show that such a
solution is indeed a critical solution, one should investigate
near-critical configurations to see whether they describe black hole
formation. However, as a strictly negative cosmological constant is necessary
for the existence of black holes in 2+1 dimensions \cite{BTZ}, one
should first extend the Garfinkle solutions to $\L < 0$, a hard
problem which has not yet been solved in generality. 

In this talk, we shall present a new class of $\L = 0$ CSS solutions
which can be extended to $\L < 0$, and show that a class of nearby solutions
do indeed describe black hole formation. After reviewing the Garfinkle
solutions, we describe the new CSS solutions and their extension
to $\L \neq 0$ quasi-CSS solutions. We then perform the linear
perturbation analysis in the background of these threshold
solutions. Imposing appropriate boundary conditions, we obtain discrete 
modes describing black hole formation, and discuss the determination of 
the critical exponent.

\section{CSS and quasi-CSS solutions}

The Einstein-massless scalar field equations with cosmological
constant are
\be
G_{\mu\nu} - \L g_{\mu\nu} = \part_{\mu}\phi\part_{\nu}\phi - 
(1/2) g_{\mu\nu}\part^{\lambda}\phi\part_{\lambda}\phi, \qquad
\nabla^2\phi = 0.
\ee 
Assuming rotational symmetry, we make the double null ansatz: 
\be\lb{an} 
ds^2 = e^{2\s(u,v)}dudv -r^2(u,v)d\theta^2, \quad \phi = \phi(u,v).
\ee
The field equations then take the form
\ba 
& r_{,uv} = (\L/2) r e^{2\s}, & \lb{ein1} \\ & 2\s_{,uv} = (\L/2)
e^{2\s} - \phi_{,u}\phi_{,v}, & \lb{ein2} \\ & 2\s_{,u} r_{,u}
-r_{,uu} = r\phi_{,u}^2, & \lb{ein3} \\ & 2\s_{,v} r_{,v}  -r_{,vv} =
r\phi_{,v}^2, & \lb{ein4}\\ & 2r\phi_{,uv} + r_{,u}\phi_{,v} +
r_{,v}\phi_{,u} = 0 & \lb{phi}\,.  
\ea
Assuming $\L = 0$, Garfinkle found \cite{gar} the following exact CSS 
solutions 
\ba
ds^2 & = & -A\left(\frac{(\sqrt{\hat{v}} +
\sqrt{-\hat{u}})^4}{-\hat{u}\hat{v}}\right)^{c^2}\,d\hat{u}\,d\hat{v}
- (\hat{v}+\hat{u})^2\,d\theta^2\,, \lb{garm} \\ \phi & = &
-2c\ln(\sqrt{\hat{v}} + \sqrt{-\hat{u}})\,, \lb{garp}
\ea 
depending on an
arbitrary constant $c$ and a scale $A > 0$. These solutions are
continuously self-similar with homothetic vector
$(\hat{u}\part_{\hat{u}} + \hat{v}\part_{\hat{v}})$ and, for $c \simeq
1$, agree well near the singularity with numerical results of
\cite{CP}. However, it has not been possible up to now to extend them
to solutions of the full $\L < 0$ equations with both CSS behaviour
near the singularity, and the correct AdS behaviour at spatial infinity. 

From the Garfinkle class of CSS solutions (\ref{garm})-(\ref{garp}), we derive
a new class of CSS solutions by an infinite
boost $\hat{u} \to \lambda^{-1} \hat{u}$, $\hat{v} \to \lambda\hat{v}$ 
($\lambda \to 0$), 
arriving at 
\be\lb{new1}
ds^2 = -\bar{A}(-\hat{u}/\hat{v})^{c^2}d\hat{u}d\hat{v} - \hat{u}^2 d\theta^2.
\ee
These new solutions obviously inherit the CSS behaviour of the
original Garfinkle solutions. The transformation to new null
coordinates $u = - (-\hat{u})^{1+c^2}$, $v = \hat{v}^{1-c^2}$ leads to the
simpler form of these solutions
\be\lb{new2} 
ds^2 = dudv - (-u)^{2\alpha} d\theta^2, \quad \phi = -c\alpha\ln(-u)  
\ee
($\alpha = 1/(1+c^2)$), which may be checked to solve exactly Eqs. 
(\ref{ein1})-(\ref{phi}). 

The generic solution (\ref{new2}) has four Killing vectors. While this
spacetime is devoid of scalar curvature singularities, the study of 
geodesic motion shows that for $c^2 < 1$ only radial geodesics can be 
continued through the null 
line $u = 0$, while nonradial geodesics terminate at the singular
point $u = 0, v \to +\infty$. On the other hand, for $c^2 > 1$
nonradial geodesics terminate on the null line  $u = 0$. So the $c^2 =
1$ solution is an extreme solution separating the $c^2 \le 1$
solutions with a point singularity from the $c^2 > 1$ solutions with a
null line singularity.

Now we proceed to extend the second class of $\L = 0$ CSS
solutions (\ref{new2}) to exact quasi-CSS solutions of the full $\L < 0$
equations (the self-similarity being then broken by the cosmological
constant). We make the ansatz 
\be\lb{anext} 
ds^2 = e^{2\nu(x)}dudv - (-u)^{2\alpha}\rho^2(x)d\theta^2, \quad \phi
= -c\alpha\ln(-u) + \psi(x) \quad (x \equiv uv). 
\ee 
Inserting this
ansatz into the field equations (\ref{ein1})-(\ref{phi}) leads to the system 
\ba 
x\rho'' + (1+\alpha)\rho' & = & (\L/2)\rho
e^{2\nu}, \lb{einx1}\\ 2(x\nu'' + \nu') +\psi'(x\psi'-c\alpha)
& = & (\L/2)e^{2\nu}, \lb{einx2}\\ x^2(-\rho'' + 2\rho'\nu'-
\rho\psi'^2) + 2\alpha x(- \rho' + \rho(\nu'+c\psi')) & = & 0
\lb{einx3} \\ - \rho'' + 2\rho'\nu'- \rho\psi'^2 & = & 0 \lb{einx4} \\
2x(\rho\psi')' + (2+\alpha)\rho\psi' & = & c\alpha\rho'. \lb{phix} 
\ea A simple first integral is
\be\lb{rhospsi} 
\rho = e^{\nu+c\psi}.  
\ee

The numerical solution of this system with the boundary conditions 
\be\lb{boundx} \rho(0) = 1, \quad \nu(0) = 0, \quad \psi(0) = 0 
\ee 
leads \cite{crit} to a unique extension which reduces  to
(\ref{new2}) near the singularity $u = 0$ and which, for $x \to x_1 < 0$, is
asymptotic to the $AdS_3$  (anti-de-Sitter) spacetime
\be
ds^2 = (X^2 + 1) dT^2 - \frac{dX^2}{X^2+1} - X^2d\theta^2.
\ee

\section{Perturbations}

To ascertain whether the quasi-CSS solutions of the preceding section
are threshold solutions for black hole formation, we study their
linear perturbations, along the lines of the analysis of \cite{fro}. 
The relevant time parameter in critical collapse being
the retarded time \cite{gar} $T = -\ln(-U) = -
\alpha\ln(-u)$, we expand these perturbations in modes
proportional to $e^{kT} = (-u)^{-k\alpha}$. Keeping only one
mode, we decompose the perturbed fields as 
\ba
r & = & (-u)^{\alpha}[\rho(x) + \lambda(-u)^{-k\alpha}\tr(x)], \lb{pr} \\
\s & = & \nu(x) + \lambda(-u)^{-k\alpha}\ts(x), \lb{ps}
\\ \phi & = & -c\alpha\ln(-u) + \psi(x) +
\lambda(-u)^{-k\alpha}\tp(x). \lb{pp} 
\ea 

The linearization of the Einstein equations (\ref{ein1})-(\ref{phi}) in
the small parameter $\lambda$ leads to the fourth order system for the 
perturbations $\tr$, $\ts$, $\tp$:
\ba 
& & x\tr'' + (1+\alpha-k\alpha)\tr' = (\L/2)\, e^{2\nu}(\tr +
2\rho\ts), \lb{pein1}\\ & & 2x\ts'' + 2(1-k\alpha)\ts' =
\L e^{2\nu}\ts - (2x\psi'-c\alpha)\tp' +
k\alpha\psi'\tp, \lb{pein2}\\ & & -(1-k)x\tr' +
((1-k)x\nu'+ (k/2)(2\alpha-1-k\alpha))\tr + \rho x\ts' -
k(x\rho' + \alpha\rho)\ts =  \nonumber \\ & &
\qquad \qquad \qquad -\rho(cx\tp' -
k(c\alpha-x\psi')\tp) - cx\psi'\tr,
\lb{pein3}\\ & & 2(\rho'\ts' + \nu'\tr') - \tr'' = \psi'(2\rho\tp' +
\psi'\tr), \lb{pein4}\\ & & 2x\rho\tp'' + (2x\rho' +
(2+\alpha-2k\alpha\rho)\tp' - k\alpha\rho'\tp
+ (2x\psi'- c\alpha)\tr' \nonumber \\ & & \qquad
\qquad \qquad + (2x\psi'' +
(2+\alpha-k\alpha)\psi')\tr = 0.  \lb{pphi} 
\ea
This system admits the exact, spurious solutions (gauge modes)
\ba 
& \tr_k^{(1)} = (-x)^{p_+}\rho', \;  
\ts_k^{(1)} = (-x)^{p_+}\nu'-\frac{p_+}2(-x)^{p_+-1}, \;
\tp_k^{(1)} = (-x)^{p_+}\psi', \\ 
& \tr_k^{(2)} = x\rho'(x) + \alpha\rho, \; 
\ts_k^{(2)} = x\nu'+\frac{p_-}2, \;
\tp_k^{(2)} = x\psi'- c\alpha,
\ea
($p_{\pm} = 1 \pm \ka$) generated respectively by the small gauge 
transformations $v \to v -
\lambda v^{p_+}$ and $u \to u - \lambda (-u)^{p_-}$.
So, up to gauge transformations, the
general solution of this system depends only on two integration
constants, which should be determined by enforcing appropriate
boundary conditions.

The three boundaries of our quasi-CSS solutions are $u =0$, $v = 0$
(or $x = 0$), and $x = x_1$ (the AdS boundary). On the singular
boundary $u = 0$, it seems natural to impose that the perturbed metric
component $r$ does not diverge too quickly, which would conflict with
the linear approximation. However we shall argue that it may be
preferable to replace this condition (which is subject to some
ambiguity) with a boundary condition on the apparent horizon.  On the
second boundary $v = 0$, we shall require that  the perturbed solution
matches smoothly the original quasi-CSS solution  \be\lb{bound2}
\tr(0) = 0, \quad \ts(0) = 0, \quad \tp(0) = 0,  \ee and moreover that
the perturbations be (as the unperturbed quasi-CSS solution) analytic
in $v$. This last condition ensures regularity on the line $v = 0$. 
Finally, the  perturbations should grow slowly (i.e. not
faster than the original fields) as the AdS boundary $x = x_1$ is
approached. It turns out that this last condition is always satisfied 
up to gauge transformations. This is due to the fact that the
perturbed scalar field becomes negligible at spatial infinity, so that 
the asymptotic behavior is that of $AdS_3$ (the effect of gauge
transformations is to shift the value $x_1$ for which the AdS boundary
is attained). 

To enforce our boundary conditions on the line $v = 0$, it is enough to
know the behavior of the solutions of the linearized system near $x =
0$. Up to gauge transformations, this is given in terms of two
integration constants $B$, $C$ by 
\ba 
& &\tr(x) \simeq B(-x) + \L C(-x)^{1+(k-1/2)\alpha},\lb{rti} \\ 
& &\ts(x) \simeq -\tp(x) \simeq - (1+\alpha-k\alpha)\L^{-1}B +O(x) 
\nonumber \\ & & \qquad \qquad - 
(1+\alpha/2)(1+(k-1/2)\alpha)C(-x)^{(k-1/2)\alpha}. \lb{sti}
\ea for $k \neq 1/2$ (in the degenerate case $k = 1/2$, the power law
$(-x)^{(k-1/2)\alpha}$ is replaced by a logarithm).
From these behaviors, we find that the smooth matching 
conditions at $v = 0$ are satisfied if either
\be
C = 0, \quad k = c^2 + 2, \lb{solb} 
\ee
(in this case the analyticity condition is also satisfied), or $B = 0$
which, taking into account the analyticity condition, implies
\be 
B = 0, \quad k = n(c^2+1)+1/2, \lb{sola} 
\ee
where $n$ is a positive integer. In this last case the boundary
condition that the perturbed $r$ does not diverge too fast (that is,
no faster than in the BTZ case, see Sect. 4) when $u \to 0$ with $x$
fixed selects the eigenvalue $n = 1$, i.e. $k = c^2+3/2$.

\section{Black hole formation}

Black hole formation is characterized by the appearance of a central
singularity hidden by an apparent horizon. The definition of these
notions can be quite ambiguous in linearized perturbation theory. For
our present purpose we shall identify the central singularity with
the coordinate singularity of the linearized perturbed metric 
\be\lb{csing}
\sqrt{g} = e^{2\s}r = (-u)^{\a}e^{2\nu(x)}[\rho(x) +
(-u)^{-\ka}(\tr(x) + 2\rho(x)\ts(x))] = 0,  
\ee 
and define the apparent horizon by
\be\lb{aphor} 
r_{,v} = (-u)^{1+\a}(\rho' + (-u)^{-\ka}\tr') = 0. 
\ee

\begin{figure}
\hspace{40pt}\epsfxsize=100pt\epsfbox{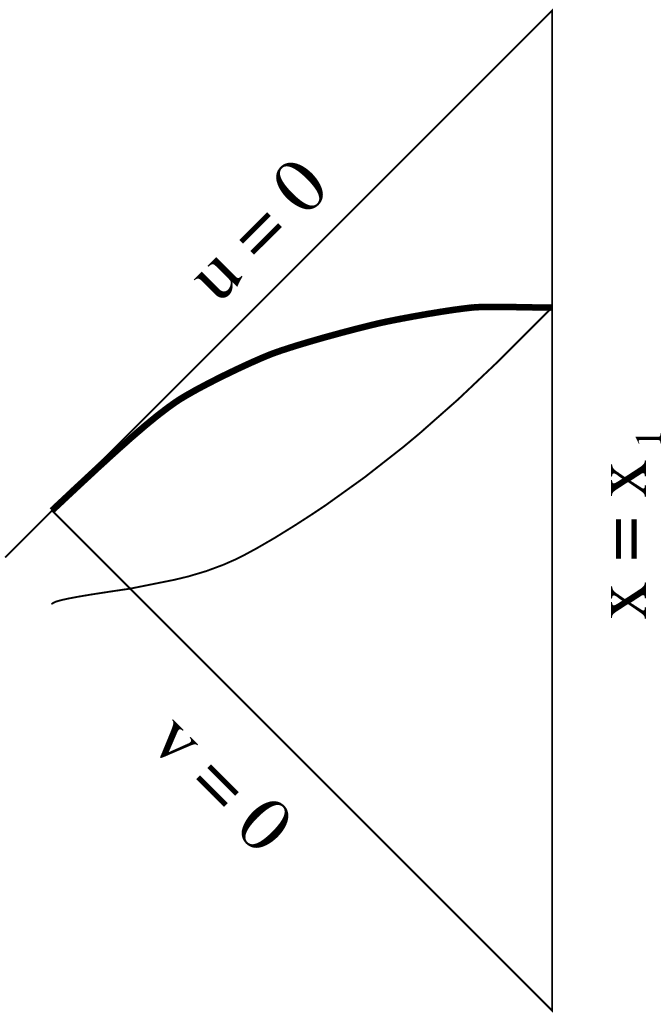}\hfill\epsfxsize=100pt
\epsfbox{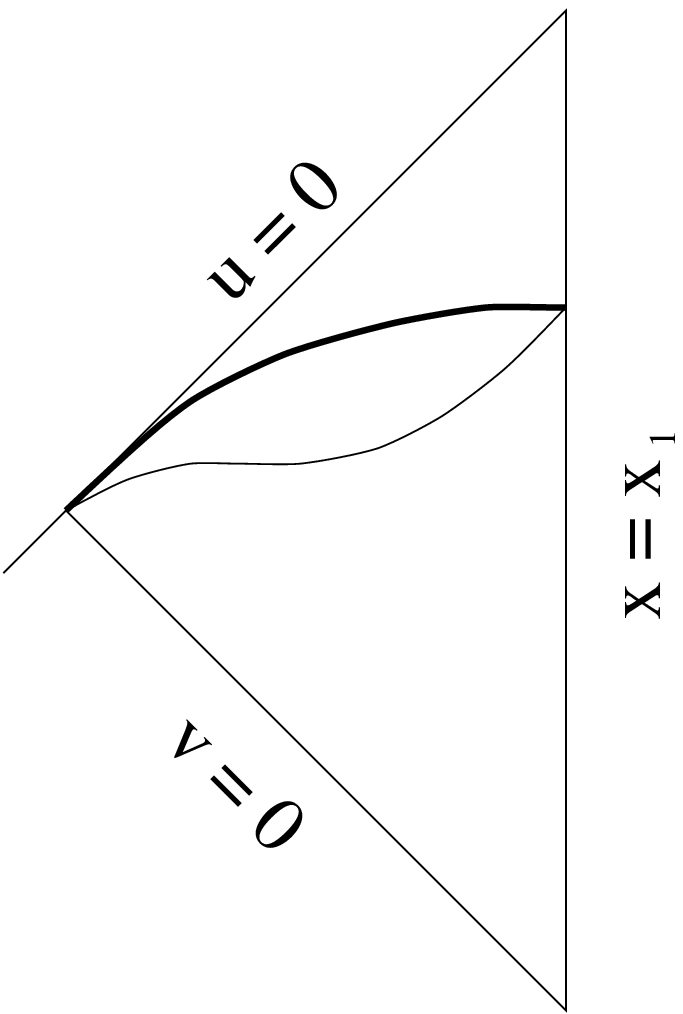}\hspace{40pt}
\caption{Perturbative singularity (thick curve) and apparent horizon
for  solutions (\ref{solb}) (left) and (\ref{sola}) (right)}
\end{figure}

In the case of the solution (\ref{solb}), for $\lambda B < 0$, a spacelike
coordinate singularity appears at $v = 0$, hidden behind a preexisting
spacelike apparent horizon, which seems to be eternal (Fig. 1). Such a
situation does not seem
to correspond to collapse, but rather to a static configuration. 
Indeed, consider the exact static solution of the Einstein equations
(\ref{ein1})-(\ref{phi}) with $\L = 0$ \cite{sig}
\be\lb{stat0}
ds^2 = A r^{c^2}(dt^2-dr^2)-r^2d\theta^2, \qquad \phi = c\ln r.
\ee
Putting $r = (-u)^{\a}+\lambda v$, $t = -(-u)^{\a}+\lambda v$, and 
expanding in powers of the small parameter $\lambda$, we get to first order
\ba
ds_0^2 & = &
\bar{A}(1-\lambda c^2(-u)^{-\a-1})dudv-(-u)^{2\a}
(1-\lambda c^2(-u)^{-\a-1})^2d\theta^2, \\ 
\phi & = & -c\ln(-u) + \lambda c(-u)^{-\a-1}x,
\ea  
which agrees with the first order expansions (\ref{pr})-(\ref{pp}) for
$\L = 0$ and $k\alpha = \alpha+1$, i.e. $k = c^2+2$. We conjecture that   
the full $\L < 0$ perturbed solution with $k = c^2+2$ results from a
similar expansion of the ``non-topological soliton'' singular static solutions
(generalizing (\ref{stat0}) to $\L < 0$) \cite{sigcosm}
\ba ds^2 & =
& A\,|\rho - \rho_+|^{1/2+a}\,|\rho - \rho_-|^{1/2-a}\,dt^2 -
\,\frac{4|\L|}{A}\,|\rho - \rho_+|^{1/2-a}\,|\rho - \rho_-|^{1/2+a}\,
d\theta^2 \nonumber \\ & & + \,\frac{d\rho^2}{4\L(\rho - \rho_+)(\rho
- \rho-)}\,, \qquad \phi = \sqrt{\frac{1-4a^2}8}\,
\ln\left(\frac{|\rho-\rho_+|}{|\rho-\rho_-|}\right)
\label{cicca}
\ea 
(with $2a = (c^2-2)/(c^2+2)$).
This conjecture is certainly true in the sourceless case $c = 0$, for
which these solutions reduce to the BTZ black holes 
\be\lb{btz1} 
ds^2 = (r^2/l^2-M)dt^2 - \frac{dr^2}{(r^2/l^2-M)} -
r^2d\theta^2, \qquad \phi = 0 
\ee 
($l^2 = -\L^{-1}$). The BTZ metric can be rewritten in the double-null
form \cite{crit}
\be
ds^2 = \frac1{(1+uv/4l^2)^2}[dudv - (u+(M/4)v)^2d\theta^2],
\ee
which corresponds to the first order expansion
(\ref{pr})-(\ref{pp}) for $\rho = e^{\nu} = (1+uv/4l^2)^{-1}$, $\phi = 0$,
$\lambda = M/4$, $k = 2$ (the linearized solution is exact in this case).

Consider now the second solution (\ref{solb}). For $C > 0$, a
coordinate singularity and an apparent horizon, both spacelike, appear 
simultaneously at $v = 0$, and converge towards the AdS boundary $x =
x_1$ for a finite value $r_{AH}$ of the horizon radius (Fig. 1). 
The fact that this apparent horizon is born together
with the singularity suggests that it is not automatically
regular. Indeed, the curvature invariant 
\be
R + 6\L = 4e^{-2\s}\phi_{,u} \phi_{,v},
\ee
evaluated on the apparent horizon, generically diverges on the
original null singularity $u = 0$ because $\phi_{,v}|_{AH}\propto x^{-1}
(x\to0)$. So the apparent horizon can be regular at birth only if  
\be\lb{regb}
e^{-2\s}\phi_{,u}|_{AH}(x=0) = 0,  
\ee
which leads to an eigenvalue equation relating $n$ and $c^2$. While
there is some ambiguity in the linearization of (\ref{regb}), it seems
to lead \cite{crit} to a single solution $n = 1$, $c^2 \simeq 1$.

Near-critical collapse is characterized by a critical exponent
$\gamma$, defined by the scaling relation 
\be
Q \propto |p - p^*|^{s\gamma}
\ee
for a quantity $Q$ with dimension $s$ depending
on a parameter $p$ (with $p = p^*$ for the critical solution).
In the perturbative approach, the choice $Q= r_{AH}$ (the apparent 
horizon radius, $s = 1$) and
$p - p^* = \lambda$ (the perturbation amplitude) leads to $\gamma =
1/k$ \cite{fro}. For the static BTZ black hole (solution (\ref{solb}) with
$c^2 = 0$), we thus obtain 
\be
\gamma_{(BTZ)} = 1/2,
\ee 
in agreement with previous
analyses. In the case of genuine scalar field collapse (solution
(\ref{sola}) with $n=1$) we obtain for the preferred value $c^2 = 1$
the critical exponent 
\be
\gamma = 2/5.
\ee 
This differs from the
two conflicting values ($\gamma \sim 1.2$ and $\gamma \sim 0.8$) found
in the numerical analyses of \cite{CP}, but is of the same order of magnitude. 

\section{Conclusion}

We have constructed a one-parameter ($c^2$) family of exact time-dependent 
solutions to the
Einstein equations with a negative cosmological constant and a
massless scalar field as source. These solutions are continuously
self-similar near the singularity, and asymptotically AdS at spatial
infinity. We have then analyzed the linear perturbations of these
solutions, and discussed the subtle question of boundary conditions. 
These led to two possible growing modes. The first mode,
(\ref{solb}), is characterized by the presence of a presumably eternal
apparent horizon. This is simply the linearization of the static BTZ 
black hole in the sourceless case, and very probably of static
singular solutions in the general coupled case. The second mode,
(\ref{sola}), exhibits the characteristic features of black
hole formation, i.e. the simultaneous appearance of a spacelike
singularity and its shielding apparent horizon, which then expands up
to a finite size. The condition of regularity at birth of this
apparent horizon selects a unique value of the parameter $c^2$, leading to    
a value of the critical exponent which is of the order of the values
derived from the numerical simulations. While these results are rather
satisfactory, we think that further analytical work on this simple 2+1
dimensional model should further clarify the situation and help us to 
understand better the phenomenon of black hole formation.

\bigskip
We would like to thank Marco Cavagli\`a for discussions.

\end{document}